# Singular Short Range Potentials in the J-Matrix Approach

M. S. Abdelmonem, I. Nasser, and H. Bahlouli
*Physics Department, King Fahd University of Petroleum & Minerals, Dhahran 31261, Saudi Arabia*

U. Al-Khawaja
*Physics Department, United Arab Emirates University, Al-Ain, UAE.*

A. D. Alhaidari
*Shura Council, Riyadh 11212, Saudi Arabia*

We use the tools of the J-matrix method to evaluate the S-matrix and then deduce the bound and resonance states energies for singular screened Coulomb potentials, both analytic and piecewise differentiable. The J-matrix approach allows us to absorb the 1/r singularity of the potential in the reference Hamiltonian, which is then handled analytically. The calculation is performed using an infinite square integrable basis that supports a tridiagonal matrix representation for the reference Hamiltonian. The remaining part of the potential, which is bound and regular everywhere, is treated by an efficient numerical scheme in a suitable basis using Gauss quadrature approximation. To exhibit the power of our approach we have considered the most delicate region close to the bound-unbound transition and compared our results favorably with available numerical data.

Singular short range potentials, such as the Yukawa [1] and Hulthén potentials [2], to mention only a few, are used in various areas of physics to model singular but short-range interactions. In high energy physics, for example, they are used to model the interaction of hadrons in short range gauge theories where coupling is mediated by the exchange of a massive scalar meson [1]. In atomic and molecular physics, it represents a screened Coulomb potential due to the cloud of electronic charges around the nucleus, which could be treated in the Thomas-Fermi approximation [3]. These potentials are also used to describe the shielding effect of ions embedded in plasmas where it is called the Debye-Hückel potential [4]. It has also been used to describe the interaction between charged particles in plasmas, solids and colloidal suspensions [5]. They were also used to describe shallow impurities [6] whose spectra are usually well understood in terms of the effective mass approximation. Usually short range potentials are defined only in terms of two basic parameters, namely the strength of the potential, and the range of the potential. The generic form of these potentials can be written as follows

$$V(r) = -\frac{A}{r} F(\mu r) \qquad (1)$$

where $A$ and $1/\mu$ are the strength and range of the potential, respectively. We will assume that $F(\mu r)$ is bounded everywhere with a limit equal to unity as $r \to 0$ [ i.e. $F(0) = 1$] so that $V(r)$ has a Coulomb-like behavior at small distances. In addition, we impose that $F(r)$ is a fast decaying function at large distances, which results in the short-range behavior of the potential; however, it is not required to be analytic. Due to the singular behavior of these potentials at the origin, the computations of the resonances are a little delicate and hence this subject did not receive adequate attention in the literature [7]. The solution of the Schrödinger equation for these potentials has been investigated extensively in the past using various numerical and perturbative approaches since exact analytical solutions are not possible (very few are known only for S-wave) [8]. Despite the short-range behavior of the potential due to the decaying envelop function $F(r)$, the $r^{-1}$ singularity at the origin makes the task of obtaining stable and accurate solutions a non-trivial task. Contrary to the traditional belief that the 1/r singularity of the potential cannot be seen experimentally due to the finite size of the nucleus [9], it has been show in recent year that higher order harmonics generation from atoms and molecules put an intense laser field are very sensitive to the Coulomb singularity [10].

Most of the perturbative and variational calculations found in the literature suffer from the limited accuracy when considering a wider range of potential parameters. Our approach constitutes a significant contribution in this regard. It is inspired by the J-matrix method [11] that handles this particular singularity not just accurately but in fact, exactly leaving the remaining non-singular and finite part to be easily treated numerically to the desired accuracy.

The J-matrix method is an algebraic method for extracting resonance and bound states information using computational tools devised entirely in square integrable bases [11]. In this approach the total Hamiltonian is split in two parts: a reference Hamiltonian $H_0$ and the remaining terms which are combined into an effective potential $U(r)$. The reference Hamiltonian contains the part of the potential which belongs to the class of exactly solvable problems that could include singular interactions like $r^{-1}$ and $r^{-2}$ [11]. However, the effective potential $U(r)$ will be treated numerically. Therefore, for meaningful results, it must be non-singular, bounded everywhere and, preferably but not necessarily, short-range [12]. The real power of our approach comes from the unique feature of J-matrix method that allowed us to isolate the $r^{-1}$ singularity of the singular potential and absorb it into the reference Hamiltonian where it is treated analytically.

The analytic continuation of the S-matrix to the second sheet in the complex energy plane in combination with complex scaling provides the most rigorous basis for the mathematical definition of resonances as being the poles of the S-matrix. Thus in our present work we adopt this approach to compute the resonances. These pole are written as $E = E_R \pm i E_I$ where the resonance position is $E_R$ and its width is given by $\Gamma = 2E_I$. There are many techniques that enable us to evaluate the S-matrix. One such approach is to use the Jost function and its analytic properties [13]. In another approach, Yamani and Abdelmonem [14] showed how to calculate the S-matrix at the real Harris energy eigenvalues and, subsequently perform analytic continuation of the S-matrix to the complex energy plane. In our present work, we will evaluate the resonance and bound states energies associated with the singular short range potential by combining the properties of the S-matrix, the complex rotation method [15] and the analytical and computational power of the J-matrix approach [11]. The unique properties of the latter will enhance accuracy and improve efficiency in locating the resonance positions and widths without extending too much the computing time or reducing numerical stability. It will enable us to include, without truncation, all of the exactly solvable part of the Hamiltonian as an infinite tridiagonal matrix (usually referred to as the $H_0$ problem in the J-matrix literature). Recently the S-matrix has been calculated using the J-matrix approach for the Morse and inverse Morse potentials [16].

The time-independent Schrödinger equation for a particle of mass $m$ and charge $q$ in the combined field generated by the Coulomb potential and a spherically symmetric potential $V(r)$ reads as follows

$$(H-E)|\psi\rangle = \left[-\frac{1}{2}\frac{d^2}{dr^2} + \frac{\ell(\ell+1)}{2r^2} + \frac{Z}{r} + V(r) - E\right]|\psi\rangle = 0, \quad (2)$$

where we have used the atomic units $\hbar = m = q = 1$ and length is measured in units of $a_0 = 4\pi\epsilon_0\hbar^2/mq^2$. Because the $r^{-1}$-type singularity of the potential is easily handled by the J-matrix method [11], we absorb it in the reference Hamiltonian by writing it as follows

$$H_0 = -\frac{1}{2}\frac{d^2}{dr^2} + \frac{\ell(\ell+1)}{2r^2} + \frac{\mathcal{Z}}{r}, \quad (3)$$

where $\mathcal{Z} = Z - A$. Therefore, the effective potential $U = H - H_0$, which will be treated numerically, has the following form [17]

$$U(r) = \frac{A}{r}[1 - F(\mu r)]. \quad (4)$$

One can easily verify that this effective potential is regular and bounded everywhere. Consequently, we can evaluate its contribution (matrix elements) in a suitable $L^2$ basis to the desired accuracy using any preferred numerical integration scheme. The $r^{-1}$-type singularity in the reference Hamiltonian and the requirement that its matrix representation be tridiagonal dictate that the J-matrix basis used should be the "Laguerre basis" defined as [18]

$$\phi_n(x) = a_n x^\alpha e^{-x/2} L_n^\nu(x); \qquad n = 0, 1, 2, .. \quad (5)$$

where $x = \lambda r$, $\lambda > 0$, $\alpha > 0$, $\nu > -1$, $L_n^\nu(x)$ is the Laguerre polynomial, and $a_n$ is the normalization constant $\sqrt{\lambda \Gamma(n+1)/\Gamma(n+\nu+1)}$. Choosing $\alpha = \ell + 1$ and $\nu = 2\ell + 1$ gives the following tridiagonal matrix representation for $H_0$ [19]

$$\frac{8}{\lambda^2}(H_0)_{nm} = \left(2n + \nu + 1 + \frac{8\mathcal{Z}}{\lambda}\right)\delta_{n,m} + \sqrt{n(n+\nu)}\delta_{n,m+1} + \sqrt{(n+1)(n+\nu+1)}\delta_{n,m-1}. \quad (6)$$

Now, the only remaining quantity that is needed to perform the calculation is the matrix elements of the effective potential $U(r)$. This is obtained by evaluating the integral

$$U_{nm} = \int_0^\infty \phi_n(\lambda r) U(r) \phi_m(\lambda r) dr = \lambda^{-1} a_n a_m \int_0^\infty x^\nu e^{-x} L_n^\nu(x) L_m^\nu(x) [xU(x/\lambda)] dx. \quad (7)$$

The evaluation of such an integral for a general effective potential is almost always done numerically. We use the Gauss quadrature integral approximation [20], which gives

$$U_{nm} \cong \sum_{k=0}^{N-1} \Lambda_{nk} \Lambda_{mk} [\varepsilon_k U(\varepsilon_k/\lambda)], \quad (8)$$

for adequately large integer $N$. $\varepsilon_k$ and $\{\Lambda_{nk}\}_{n=0}^{N-1}$ are the $N$ eigenvalues and corresponding normalized eigenvectors of the $N \times N$ tridiagonal basis overlap matrix $\langle \phi_n | \phi_m \rangle$. Therefore, the reference Hamiltonian $H_0$ in this representation, which is given by Eq. (6), is accounted for in full. On the other hand, the effective potential $U$ is approximated by its matrix elements in a subset of the basis. In this letter, we limit our investigation to the Coulomb-free interaction with $Z = 0$. Even though the effective potential has a long range Coulomb part, our numerical approach selects the right space dimension, N, so as to stabilize our calculation and ensure that such a truncation does not affect the potential matrix elements and hence accounts correctly for the long range part of the effective potential.

Our approach in finding resonance and bound state energies makes use the "direct method" based on the J-matrix calculation of the scattering matrix in the complex energy plane. Bound states are associated with negative real poles of the S-matrix while resonances are associated with the complex poles that have positive real parts and negative imaginary parts. The $N^{th}$ order S-matrix in the J-matrix approach, is defined by [14]

$$S(E) = T_{N-1}(E) \frac{1 + g_{N-1,N-1}(E) J_{N-1,N}(E) R_N^-(E)}{1 + g_{N-1,N-1}(E) J_{N-1,N}(E) R_N^+(E)}, \qquad (9a)$$

Where

$$g_{N-1,N-1}(z) = \frac{1}{N+2\ell+1} \sum_{n=0}^{N-1} \frac{\Lambda_{N-1,n}^2}{\varepsilon_n - z}$$

$$= \frac{1}{N+2\ell+1} \left[ \prod_{m=0}^{N-2} (\tilde{\varepsilon}_m - z) \Big/ \prod_{n=0}^{N-1} (\varepsilon_n - z) \right] \qquad \text{, and} \qquad (9b)$$

and

$$T_n = \frac{c_n - is_n}{c_n + is_n} \; ; \quad R_n^\pm = \frac{c_n \pm is_n}{c_{n-1} \pm is_{n-1}} . \qquad (9c)$$

$g_{nm}(z)$ is the inverse of the matrix $(H - z)$ where $H$ is the $N \times N$ finite matrix given by the first line in (6). The eigenvalues and the corresponding normalized eigenvectors of the finite matrix $H$ are denoted by $\{\varepsilon_n\}_{n=0}^{N-1}$ and $\{\Lambda_{nm}\}_{n,m=0}^{N-1}$, respectively. $\{\tilde{\varepsilon}_n\}_{n=0}^{N-2}$ are the eigenvalues of the truncated $H$ obtained by removing the last row and last column. The matrix representation of the reference wave operator, whose elements are defined by $J_{m,n}(E) = \langle \phi_m | (H_0 - E) | \phi_n \rangle$, is tridiagonal and symmetric. The quantities $s_n$ and $c_n$ are the expansion coefficients of the two independent asymptotic solutions of the reference wave equation $(H_0 - E)\psi = 0$ which, in the J-matrix terminology, are written as

$$|S\rangle = \sum_{n=0}^{\infty} s_n |\phi_n\rangle; \quad |C\rangle = \sum_{n=0}^{\infty} c_n |\phi_n\rangle, \qquad (10)$$

and are usually called the "sine-like" and "cosine-like" solutions, respectively. Table 1 gives all the elements needed to calculate the S-matrix (9a). Using the recursion relation, we can obtain $T_{N-1}(E)$ and $R_N^\pm(E)$ from $T_0(E)$ and $R_1^\pm(E)$ recursively in the form of a continued fraction as shown in [14]. Now we proceed to find the roots of $S^{-1}(E)$ using as seed the values generated by the complex rotation method [15] for different values of potential parameters and $\ell$. This strategy increases the speed of convergence of the root finding algorithm.

In our numerical implementation of the above scheme, we have selected the two most common analytic examples; the Yukawa and Hulthén potentials

$$V_Y(r) = -\frac{A}{r} e^{-\mu r} , \; V_H(r) = -A \frac{\mu}{e^{\mu r} - 1} , \qquad (11)$$

and another piecewise differentiable potential with the function $F(\mu r)$ shown in Fig. 1. The potential strength is chosen to be unity ($A = 1$) leaving us with a single parameter potential, the screening length $\mu$. This is suggested by the scaling law, which results from the original Schrödinger equation [21]:

$$E(A,\mu) = A^2 E(1, \mu/A), \quad \psi(A,\mu,r) = A^{3/2} \psi(1, \mu/A, Ar), \tag{12}$$

where $E$ and $\psi$ are the eigenvalues and eigenfunction in (2). In calculating the bound state energies, the critical value of $\mu$ [21] plays a major role in limiting the accuracy of the results. The critical screening parameter, $\mu_c$, for a particular bound state is defined as the value of the screening parameter above which the state is not to be found within the bound states spectrum. Equivalently it can also be defined as the value of the screening parameter $\mu$ at which the energy for such state is zero. For the Hulthén potential this critical value has been fitted successfully to the following analytic formula [21, 22]

$$\mu_c = \left(\frac{k}{\sqrt{2}} + 0.1654\ell + \frac{0.0983\ell}{k}\right)^{-2} \quad ; \quad k = 1, 2, 3\cdots \quad ; \quad \ell = 0, 1, 2, \cdots (n-1). \tag{13}$$

where $\ell$ is the angular momentum quantum number and k is the ordering of the eigenvalues of the bound states shifted by $\ell$ ( also called the principal quantum number in atomic physics). When the screening parameter approaches a critical value for a given state, the results become less accurate, due to instabilities caused by the absorption of the bound state into the continuum. To improve the accuracy of the numerical results in the J-matrix method, we increase the basis size, $N$, and/or decrease the basis scale parameter, $\lambda$. The later results in a better sampling in the desired region. To test this claim we did the calculation first for $\ell = 0$ where exact analytic results are available. In all our tables the energies are calculated in atomic units and results are reported only up to the precision that maintained stability and are *truncated* rather than *rounded-off*. Thus, all the results may be considered as correct up to the place they are reported. In Table 2.a we compare the bound state energies for S-states ($\ell = 0$) with exact values. One can see from this table that when the value of the screening parameter is close to $\mu_c$ the accuracy of the calculated eigenenergy is low. For example, the energy for the 3s state in this table agrees with the exact value only up to 5 decimal places. However, when we reduce the scaling parameter $\lambda$ to 0.2, we found that the agreement increased up to 10 decimal places as compared to the exact values. In the same table we show the 14s state, for which $\mu$ is very close to $\mu_c$ [23], in this case our result agrees with the exact one only up to three decimal places for $N = 50$ and $\lambda = 0.1$. To improve the accuracy of this result we needed to increase $N$ to 100 and tune $\lambda$ to its optimal value within its plateau which was found to be $\lambda = 0.06$ for which our result agree with the exact one up to eight digits. This behavior is due to the fact that for bound states very close to the continuum (near $E = 0$) the wavefunction has a long tail which leads to long range behavior; contrary to deep bound levels. Thus, the basis set should have large extensions (small length scaling, $\lambda$) to ensure that the potential is sampled correctly in regions far away from the origin. Therefore, in general, as we approach the critical screening parameter for a state, the results become less accurate. This requires a bigger basis size (large $N$) and/or a more extended basis (smaller $\lambda$) to make a better approximation of the true eigenfunctions of the Hamiltonian. In Table 2.b, we show similar results for finite angular momentum. Since the finite angular momentum case cannot be solved exactly, we have compared our results with the variational ones obtained by Stubbins [24]. We need to mention that there are many other recent references on the Hulthén potential in the literature [25] but we chose to compare our results with Stubbins [24] mainly due to the large number of significant figures of his results. The numerical parameters were chosen as $N = 50$ and $\lambda$

= 0.4, which lies in the $\lambda$ plateau between 0.2 and 0.7. From this table we see that increasing the screening parameter results in pushing the bound state with energy closest to zero (the origin in the complex energy plane) into the continuum to re-appear as a resonance. The other bound states will get closer to the origin while the resonances move away from the origin.

Similar computations were made for the Yukawa potential. However, in this case no analytic results are available for the eigenenergies or for critical values of the screening parameter; even for S-states ($\ell = 0$). Table 3 shows the bound states and resonance energies for the Yukawa potential for different values of $\ell$ and $\mu$ with $N = 50$. In this table we have chosen to be close to the critical value of the screening parameter, these values are tabulated in [26], to exhibit the bound-unbound transition of the state. Since the s-wave Yukawa potential does not support resonances the bound state will simply disappear once we cross the critical value of the screening parameter. However, for finite angular momentum we can see the bound-unbound transition occurring as we cross the critical value $\mu_c$.

We should emphasize that our approach is quite general and applies to a whole class of short-range Coulombic-type potentials, usually called screened Coulomb potentials as generalized in Eq. (1). The main characteristic of this class of potentials is that they behave like the Coulomb potential at short distances and decay exponentially at large distances. In particular, any superposition of the Yukawa and Hulthén potentials will pertain to the same class. In fact, even those piecewise differentiable potentials where the function $F(\mu r)$ in (1) is continuous but non-differentiable. Table 4 shows the result of our calculation of bound states energies for the piecewise differentiable potential with $F(\mu r)$ given by

$$F(x) = \begin{cases} x+1 & ,1 > x \geq 0 \\ 1 & ,1 \leq x \leq 2 \\ -x+4 & ,2 < x < 4 \\ 0 & , x \geq 4 \end{cases} \quad ; x = \mu r, \qquad (12)$$

and is shown in Fig. 1.a. The corresponding effective potential, $U(r)$, is shown in Fig. 1.b. In Table 4, for the case $\ell = 0$ we noticed that with increasing the screening parameter $\mu$ above the critical value $\mu_c$, the bound state (4s) disappeared. The 3s-state gets closer to the origin, while the other bound states, 1s and 2s, moved away from the origin. This is in contrast to the situation for analytic potentials where all bound states get closer to the origin coherently, see Table 2b. The same behavior is observed for higher values of $\ell$. This critical behavior requires further study and we will leave it for future investigations.

Finally, we believe that our numerical approach is more stable and rapidly convergent for a relatively small number of basis set. To test the utility of our approach we opted to calculate the bound states and resonance energies in the region close to the critical screening parameter. This usually requires more computational care because as we approach the critical screening parameter for a state, the accuracy and stability of the calculation become very demanding. This is due to the fact that bound states close to the critical transition region between bound-unbound states ($E = 0$) will have much of the wavefunction in the forbidden region [the non-classical region where $E < V(\infty)$] and are very much extended. Under these circumstances, the most suitable basis set should have long extensions (small $\lambda$) and/or a bigger size (large $N$) to ensure that the potential is sampled correctly in regions away from the origin. Having two numerical parameters at our disposal; the basis length-scale, $\lambda$, and the size of the basis set, $N$; turned out to be very efficient in fine-tuning our numerical results. Away from the critical region, our approach has already been applied to both potentials very successfully [17, 27].

**ACKNOWLEDGMENTS**

The authors acknowledge the support provided by the Physics department at King Fahd University of Petroleum & Minerals under project FT-080010. We are also grateful to "Khaled Technical & Commercial Services" (KTeCS) for the generous support.

**Table Captions:**

**Table 1:** The explicit form of the kinematic quantities $T_0(E)$, $R_1^{\pm}(E)$, and $J_{N-1,N}(E)$ in the Laguerre basis. The three-term recursion relations for $s_n$ and $c_n$ (collectively shown as $f_n$) are also given.

**Table 2:**
**(a)** Calculated bound states energies for the S-wave Hulthén potential for different values of $\mu$ compared to the exact values. **(b)** Bound states and resonance energies for the Hulthén potential for different values of $\ell$ and $\mu$. The values obtained by Stubbins [24] are given in the last column. The numerical parameters were chosen as $N = 50$ and $\lambda = 0.4$.

**Table 3:**
Bound states and resonance energies for the Yukawa potential for $N = 50$ and $\lambda = 0.3$ and different values of $\ell$ and $\mu$.

**Table 4:**
Bound states energies for the piecewise differentiable potential with $F(\mu r)$ given by Fig. 1 for different values of $\ell$ and $\mu$.[ $\lambda$=16 for $\ell$=0 and 1, $\lambda$=14 for $\ell$=2 and 3]

**Table 1**

| | |
|---|---|
| $T_0(E)$ | $e^{2i\theta} \dfrac{{}_2F_1(-\ell,1;\ell+2;e^{2i\theta})}{{}_2F_1(-\ell,1;\ell+2;e^{-2i\theta})}; \quad \cos\theta = \dfrac{8E-\lambda^2}{8E+\lambda^2}$ |
| $R_1^{\pm}(E)$ | $\dfrac{1}{(\ell+2)} e^{\mp i\theta} \dfrac{{}_2F_1(-\ell,2;\ell+3;e^{\mp i\theta})}{{}_2F_1(-\ell,1;\ell+2;e^{\mp i\theta})}$ |
| $J_{N-1,N}(E)$ | $\left(E+\lambda^2/8\right)\sqrt{N(N+2\ell+1)}$ |
| Recursion Relation | $2(\cos\theta)(n+\ell+1)f_n - \sqrt{n(n+2\ell+1)}\,f_{n-1}$ $-\sqrt{(n+1)(n+2\ell+2)}\,f_{n+1} = 0; \quad n \geq 1$ |

**Table 2.a**

| State | $\mu_c$ | $\mu$ | $N$ | $\lambda$ | $E$ (this work) | $E$ (exact) |
|---|---|---|---|---|---|---|
| 1s | 2.0 | 0.21 | 50 | 0.8 | −0.400512499 | −0.4005125 |
| 2s | 0.5 | | | | −4.205E-2 | −4.205E-2 |
| 3s | 0.22… | | | | −1.679E-4 | −1.680555...E-4 |
| 3s | | | | 0.2 | −1.6805555554E-4 | |
| 14s | 0.0102 | 0.01 | 50 | 0.10 | −1.0202E-6 | −1.02040816326530E−6 |
| 14s | | | 100 | 0.06 | −1.0204081E-6 | |

**Table 2.b**

| $\ell$ | State | $\mu_c$ [23] | $\mu$ | E (this work) | E [24] |
|---|---|---|---|---|---|
| 1 | 2p | 0.376936 | 0.18 | −4.864123176038E-2 | |
| | | | 0.20 | −4.188604921786E-2<br>5.478497896E-4 −i 3.771667228E-4 | −4.188604921786E-2 |
| | | | 0.25 | −2.661105135091E-2<br>4.453523795E-4 −i 3.3018328045E-3 | −2.661105135091E-2 |
| | 3p | 0.186486 | 0.18 | −4.7689388317E-4 | |
| 3 | 4f | 0.086405 | 0.05 | −1.0061964550933E-2 | −1.0061964550933E-2 |
| | | | 0.075 | −2.55629697807E-3<br>1.0932654251E-3 −i 6.693863637E-4 | −2.55629697807E-3 |
| | 5f | 0.059973 | 0.05 | −1.783545794710618E−3 | −1.783545794710618E−3 |
| | | | 0.10 | 2.0108248838E-3 −i 3.862579834E-4 | |
| 4 | 5g | 0.054505 | 0.05 | −1.01588159045E-3<br>8.557605324E-4 −i 3.684603746E-4 | −1.01588159045E-3 |
| | | | 0.06 | 9.563388503E-4 − i5.44931771E-5<br>1.121456108E-3 − i1.667770836E-3 | |

**Table 3**

| $\ell$ | $\mu_c$ [26] | $\mu$ | E (this work) |
|---|---|---|---|
| 0 | 1.1906 | 1.180 | −3.097E-5 |
| 1 | 0.2202 | 0.220 | −2.869723E−5 |
| | | 0.2210 | 9.81567E-5 −i 9.1777E−6 |
| 2 | 0.09135 | 9.10E-2 | −7.767498160E−5 |
| | | 9.150E-2 | 3.411464939E−5 −i 3.4952E−8 |
| 3 | 0.04983 | 4.970E-2 | −3.46170059E−5 |
| | | 4.990E-2 | 1.8018201E−5 −i 1.44E−10 |

**Table 4**

| $\ell$ | state | $\mu_c(n)$ | $\mu$ | E (this work) N=100 | E (this work) N=200 |
|---|---|---|---|---|---|
| 0 | 1s | | 0.28 | -0.779099 | -0.779097 |
| | 2s | | | -0.327726 | -0.327715 |
| | 3s | | | -0.125856 | -0.125844 |
| | 4s | 0.2827865 | | -0.0011 | -0.0028 |
| | | | | | |
| | 1s | | 0.30 | **-0.798541** | **-0.798547** |
| | 2s | | | **-0.33169** | **-0.33172** |
| | 3s | | | **-0.116940** | **-0.116948** |
| 1 | 2p | | 0.30 | -0.36829 | -0.36827 |
| | 3p | | | -0.14789 | -0.14790 |
| | 4p | 0.306178 | | -0.0049 | -0.0053 |
| | | | | | |
| | 2p | | 0.32 | -0.37700 | -0.37703 |
| | 3p | | | -0.14198 | -0.14199 |
| 2 | 3d | | 0.20 | -0.19303 | -0.19304 |
| | 4d | | | -0.08645 | -0.08644 |
| | 5d | 0.210492 | | -0.00669 | -0.00705 |
| | | | | | |
| | 3d | | 0.23 | -0.19865 | -0.19865 |
| | 4d | | | -0.07518 | -0.07517 |
| 3 | 4f | | 0.24 | -0.1063468 | -0.1063478 |
| | 5f | 0.245120 | | -0.00316 | -0.00324 |
| | | | | | |
| | 4f | | 0.26 | -0.10021 | -0.10023 |

**Figure Caption:**

**Fig. 1:**
(a) The piecewise differentiable dimensionless function $F(\mu r)$ in the potential (1) whose bound states energy spectrum is given in Table 3. (b) The corresponding effective potential, $U(r)$, in units of $\mu A$.

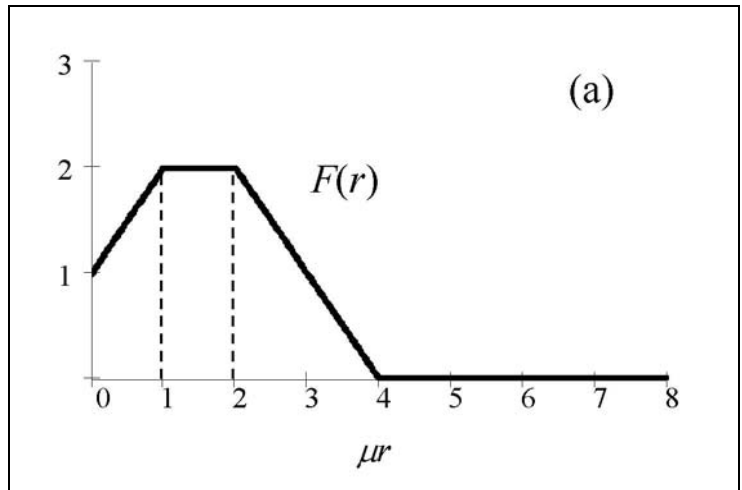

**Fig. 1.a**

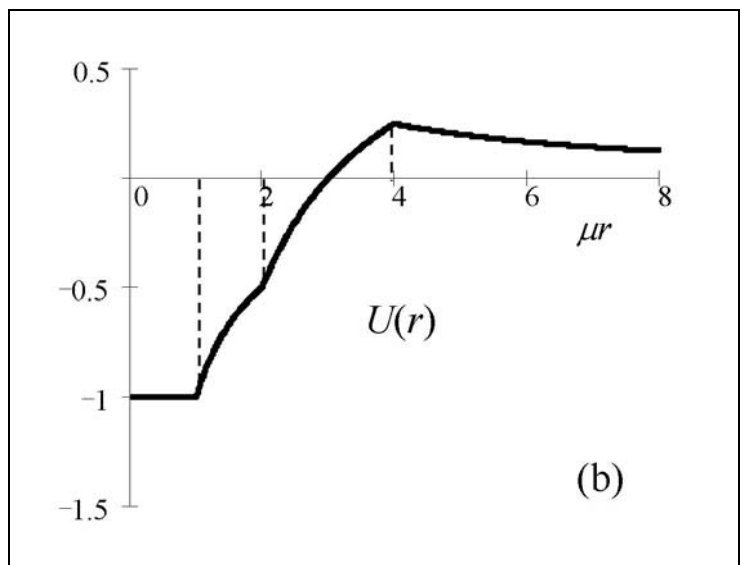

**Fig. 1.b**